\documentclass[aps,preprint,amssymb]{revtex4}
\usepackage{epsfig}

\def\xc{_{\rm xc}}

\def\rv{{\bf r}}

\def\beq{\begin{equation}}
\def\eeq{\end{equation}}

\begin{document}
\title{Study of the discontinuity of the exchange-correlation potential in an exactly soluble case}
\author{Paola Gori-Giorgi$^{1,2}$ and Andreas Savin$^1$}
\affiliation{$^1$ Laboratoire de Chimie Th\'eorique, CNRS UMR76116, Universit\'e Pierre et Marie Curie, 4 Place Jussieu, F-75252 Paris, France\\
$^2$ Afdeling Theoretische Chemie, Vrije Universiteit, De Boelelaan 1083, 1081 HV Amsterdam, The Netherlands}
\date{\today}
\begin{abstract}
It was found by Perdew, Parr, Levy, and Balduz [Phys. Rev. Lett. {\bf 49}, 1691 (1982)] and by Sham and Schl\"uter [Phys. Rev. Lett. {\bf 51}, 1884 (1983)] that the exact Kohn-Sham exchange-correlation potential of an open system may jump discontinuosly as the particle number crosses an integer, with important physical consequences. Recently, Sagvolden and Perdew [Phys. Rev. A {\bf 77}, 012517 (2008)] have analyzed the discontinuity of the exchange-correlation potential as the particle number crosses one,  with an illustration that uses a model density for the H$^-$ ion. In this work, we extend their analysis to the case in which the external potential is the simple harmonic confinement, choosing spring-constant values for which the two-electron hamiltonian has an analytic solution. This way, we can obtain the exact, analytic, exchange and correlation potentials for particle number fluctuating between zero and two, illustrating the discontinuity as the particle number crosses one without introducing any model or approximation. We also discuss exchange and correlation separately.
\end{abstract}
\pacs{boh}
\maketitle
\section{Introduction}
\label{intro}
Kohn-Sham (KS) density functional theory (DFT) (see, e.g., \cite{Koh-RMP-99}) is a successful method for electronic structure calculations, thanks to its unique combination of low computational cost and reasonable accuracy. In the Kohn-Sham formalism, the total energy of a many-electron system in the external potential $\hat{V}_{{\rm ext}}=\sum_i v_{{\rm ext}}(\rv_i)$ is rewritten as a functional of the one-electron density $\rho(\rv)$,
\beq
E[\rho]=T_s[\rho]+\int d\rv\,v_{{\rm ext}}(\rv)\,\rho(\rv)+U[\rho]+E\xc[\rho].
\label{eq_Erho}
\eeq
In Eq.~(\ref{eq_Erho}), $T_s[\rho]$ is the kinetic energy of a non-interacting system of fermions (usually called KS system) having the same one-electron density $\rho$ of the physical, interacting, system. The Hartree energy $U[\rho]$ is the classical repulsion energy, $U[\rho]=\frac{1}{2}\int d\rv\int d\rv'\rho(\rv)\rho(\rv')|\rv-\rv'|^{-1}$, and the exchange-correlation functional $E\xc[\rho]$ must be approximated. Minimization of Eq.~(\ref{eq_Erho}) with respect to the orbitals forming the KS determinant lead to the KS equations. Thus, instead of the physical problem, in KS DFT we solve the hamiltonian of a model system of non-interacting fermions in the one-body local potential $\hat{V}_{\rm KS}=\sum_i v_{{\rm KS}}(\rv_i)$, with
\begin{eqnarray}
v_{{\rm KS}}(\rv) & = & v_{\rm ext}(\rv)+v_{\rm H}(\rv)+v\xc(\rv) \label{eq_vKS}\\
v_{\rm H}(\rv) & = & \int d\rv' \frac{\rho(\rv')}{|\rv-\rv'|} \label{eq_vH}\\
v\xc(\rv) & = & \frac{\delta E\xc[\rho]}{\delta\rho(\rv)}, \label{eq_vxc}
\end{eqnarray}
and we recover the energy of the physical, interacting system, through the sum of the two functionals $U[\rho]+E\xc[\rho]$.

In Refs.~\cite{PerParLevBal-PRL-82,ShaSch-PRL-83,PerLev-PRL-83} an analysis of KS theory for systems with fluctuating particle number lead to the conclusion that the exact  exchange-correlation potential $v\xc(\rv)$ may jump discontinuously by a spatial-independent constant as the particle number crosses an integer, with important physical consequences. In the last years there has been new interest in the derivative discontinuity of $E\xc[\rho]$ (see, e.g., \cite{GruMarRub-PRB-06,GruMarRub-JCP-06,RuzPerCsoVydScu-JCP-06,Toz-JCP-03,TeaDepToz-JCP-08,MorCohYan-JCP-06,CohMorYan-SCI-08}), and its existence has been questioned in Ref.~\cite{ZahWan-PRA-04}. In a recent paper, Sagvolden and Perdew \cite{SagPer-PRA-08}  have given further support to the assumptions used to find the exchange-correlation potential discontinuity, and they have illustrated the discontinuity when the particle number crosses one, using a model density for the H$^-$ ion. They have also rigorously proved that the von Weizs\"acker functional,
\beq
T_{{\rm vW}}[\rho]=\frac{1}{2}\int d\rv |\nabla\sqrt{\rho(\rv)}|^2
\label{eq_vW}
\eeq
is the correct $T_s[\rho]$ for a system with particle number $N\le 2$.

In this work we repeat a similar analysis for electronic systems with particle number fluctuating between zero and two when the external potential is harmonic, $v_{\rm ext}(\rv)=\frac{1}{2}\omega^2 r^2$. Taut \cite{Tau-PRA-93} has shown that the corresponding hamiltonian for $N=2$ electrons is analytically soluble for some special values of $\omega$, which means that, in such special cases, we can calculate analytically the exact interacting density and the exact exchange-correlation potential \cite{FilUmrTau-JCP-94}. This way, we can illustrate the derivative discontinuity without relying on any approximation. The paper is organized as follows. In Sec.~\ref{sec_theory} we report the equations used to extract the exact exchange and correlation potentials from Taut's \cite{Tau-PRA-93} analytical solutions. The corresponding results are shown and analyzed in Sec.~\ref{sec_results}, and the last Sec.~\ref{sec_conc} is devoted to concluding remarks.

\section{Theory}
\label{sec_theory}
We consider a system with particle number fluctuating between 0 and 2, in the harmonic external potential $v_{\rm ext}(\rv)=\frac{1}{2}\omega^2 r^2$. 
This system can be thought as arising from identical quantum traps whose centers are separated by a very large distance \cite{YanZhaAye-PRL-00}. If the energy of the physical system is convex as a function of the integer particle number $M$, i.e., if $E(M)\le (E(M+1)+E(M-1))/2$, then the system with non-integer particle number $N$ is an ensemble of only the two systems with integer-particle number $M$ and $M+1$ such that $M<N<M+1$ \cite{PerParLevBal-PRL-82}. In the case treated here,  $v_{\rm ext}(\rv)=\frac{1}{2}\omega^2 r^2$, the energy of the non-interacting system is convex, and the electron-electron interaction seems to make it strictly convex (see, e.g., the results for two-dimensional harmonic traps of Refs.~\cite{RonCavBelGol-JCP-06} or the three-dimensional case treated in Ref.~\cite{TauPerCioSta-JCP-03}).
The  density of the system with particle number $0\le N\le2$, $\rho_N(\rv)=\rho_N(r)$, is then equal to \cite{PerParLevBal-PRL-82,SagPer-PRA-08}
\beq
\rho_N(\rv)=\left\{
\begin{array}{lr}
 N\rho_1(\rv) & 0\le N\le 1 \\
 (2-N)\rho_1(\rv)+(N-1)\rho_2(\rv) & 1< N \le 2
\end{array}
\right.
\label{eq_rhoN}
\eeq
The exact Kohn-Sham potential is, up to a constant, given by the functional derivative with respect to $\rho_N(\rv)$ of the von Weizs\"acker functional of Eq.~(\ref{eq_vW}),
\beq
v_{\rm KS}(r)=\frac{\nabla^2\sqrt{\rho_N(r)}}{2\sqrt{\rho_N(r)}}+{\rm const}.
\eeq
The Hartree potential $v_{\rm H}(r)$ can be easily calculated by plugging Eq.~(\ref{eq_rhoN}) into Eq.~(\ref{eq_vH}). The exchange and correlation potential $v\xc(r)$ is then calculated from Eq.~(\ref{eq_vxc}), $v\xc(r)=v_{\rm KS}(r)-v_{\rm H}(r)-v_{\rm ext}(r)$. Setting $v\xc(r\to\infty)=0$ {\em for any fixed $N$} determines the arbitrary constant in the exchange-correlation potential.

Splitting the potential $v\xc(r)$ into exchange and correlation is subtle. When $0\le N\le 1$ we know that $v_x(r)$ must exactly cancel the Hartree potential, $v_x(r)=-v_{\rm H}(r)$, and that, since the correlation energy is zero, $v_c(r)=0$. For $1\le N\le 2$ the sum $U[\rho_N]+E_x[\rho_N]$ should be equal to the expectation of the electron-electron repulsion $\hat{V}_{ee}$ over the non-interacting ensemble density matrix that yields $\rho_N(r)$. 
The non-interacting density matrix that yields $\rho_N(r)$ and corresponds to the kinetic energy functional of Eq.~(\ref{eq_vW}) is
\beq
\Gamma_0=(2-N)|\Phi_1\rangle\langle\Phi_1|+(N-1)|\Phi_2\rangle\langle\Phi_2|,
\eeq
with
\begin{eqnarray}
\Phi_1(r) & = & \sqrt{\frac{\rho_N(r)}{N}} \\	
\Phi_2(r_1,r_2) & = & \sqrt{\frac{\rho_N(r_1)}{N}}\sqrt{\frac{\rho_N(r_2)}{N}}.
\end{eqnarray}
Thus
\beq
U[\rho_N]+E_x[\rho_N]={\rm Tr}(\Gamma_0\hat{V}_{ee})=\frac{2(N-1)}{N^2}U[\rho_N].
\label{eq_tr}
\eeq
Unlike the Hartree functional $U[\rho_N]$ and the non-interacting kinetic energy functional $T_{{\rm vW}}[\rho_N]$ of Eq.~(\ref{eq_vW}), which have a simple explicit dependence on $\rho_N(\rv)$ alone, we see from Eq.~(\ref{eq_tr}) that the exchange functional $E_x[\rho_N]$ also  explicitly depends on the particle number $N$. If we take  the functional derivative $\delta/\delta\rho_N(\rv)$ of both sides of Eq.~(\ref{eq_tr}) at fixed constant $N$, the exchange potential is not discontinuous at $N=1$. If, instead, we allow little variations of $N$ with respect to $\rho_N(\rv)$, by taking into account that $N=\int d\rv \,\rho_N(\rv)$, we obtain, for the whole range $0\le N\le 2$,
\beq
v_x(r)=\left\{
\begin{array}{lr}
 -v_{\rm H}(r) & 0\le N\le 1 \\
 -\frac{(N^2-2N+2)}{N^2}\, v_{\rm H}(r)+\frac{2(2-N)}{N^3}U[\rho_N] & 1< N \le 2
\end{array}
\right.
\label{eq_vx}
\eeq
Equation~(\ref{eq_vx}) shows that, within the definition of exchange of Eq.~(\ref{eq_tr}) and allowing variations of $N$ with respect to $\rho$, the discontinuity of the exchange potential at $N=1$ is  a spatially-independent constant equal to
\beq
v_x(r)|_{N\to 1^+}-v_x(r)|_{N\to 1^-}=2 U[\rho_1].
\eeq
Exact exchange in an open system of fluctuating electron number as also been widely discussed in Ref.~\cite{PerRuzCsoVydScuStaTao-PRA-07}, where only the energy, and not the potential, has been analyzed.

When the external potential is harmonic, $v_{\rm ext}(r)=\frac{1}{2}\omega^2 r^2$, we have
$\rho_1(r)=\omega^{3/2}/\pi^{3/2}\,e^{-\omega r^2}$. 
The ground-state wavefunction of the hamiltonian with $N=2$ electrons has the form
\beq
\Psi(\rv_1,\rv_2)=\eta(R)\psi(r_{12}). \qquad R=\frac{|\rv_1+\rv_2|}{2},\;\;\;r_{12}=|\rv_2-\rv_1|.
\label{eq_Psi}
\eeq
The center-of-mass wavefunction $\eta(R)$ is a simple ground-state three-dimensional harmonic oscillator state. Taut \cite{Tau-PRA-93} has shown that for some special values of $\omega$ also the relative wavefunction $\psi(r_{12})$ has an analytic form.  For some of these special $\omega$-values we have calculated the corresponding electronic density $\rho_2(r)$,
\beq
\rho_2(r)=C\frac{e^{-\omega\,r^2}}{2\,\omega\,r}\int_0^\infty d r_{12} r_{12} t(r_{12})^2 \left(e^{-\omega (r-r_{12})^2}-e^{-\omega (r+r_{12})^2}\right),
\label{eq_rho2}
\eeq
where $t(x)=e^{\frac{1}{2}\omega\, x^2}\psi(x)$, and $C$ is a normalization constant. The function $t(x)$ is a polynomial of some finite order $n$ (depending on $\omega$), so that the corresponding $\rho_2(r)$ is completely analytical (see Appendix~\ref{app_dens}). This allows us to obtain the exact exchange-correlation potentials up to any large $r$ exactly, without introducing any approximation. The large-$r$ part of the potentials is crucial in order to illustrate the derivative discontinuity.

Notice that similar calculations of the exact exchange-correlation potential and of other DFT-related properties for the $N=2$ system have been carried out in the past, mainly aimed at comparing the exact potentials with current approximations (see e.g., \cite{FilUmrTau-JCP-94,BurPerLan-PRL-94,BurAngPer-PRA-94}). Here our aim is to calculate the exact potentials for $0\le N\le 2$ in order to study their discontinuity  as the particle number crosses one. 
\begin{figure}
\includegraphics[width=9cm]{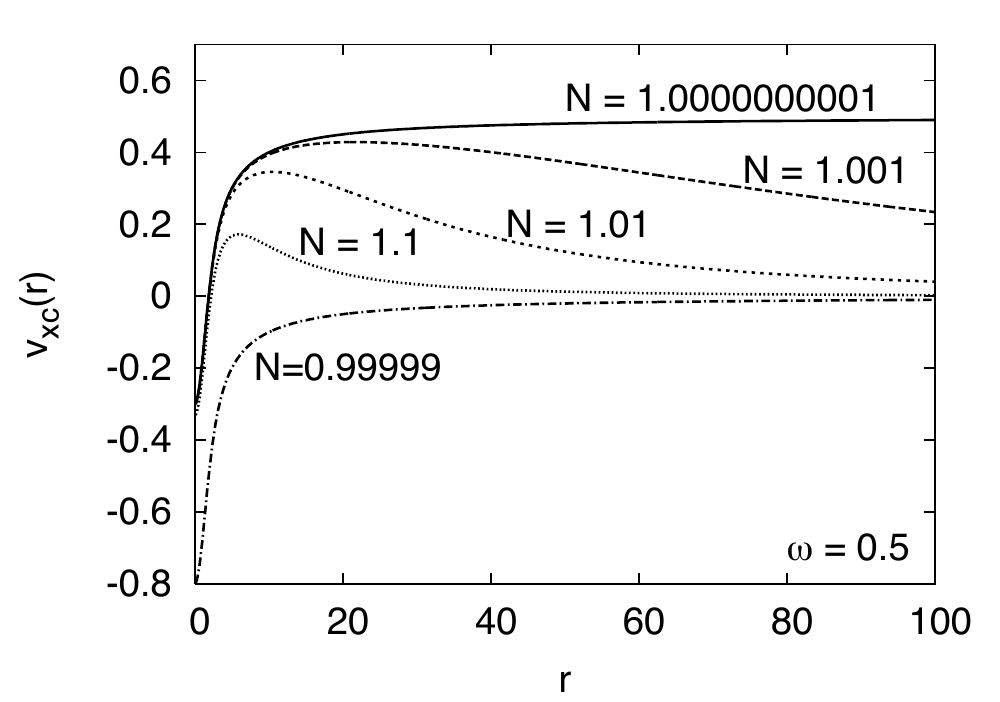}
\includegraphics[width=9cm]{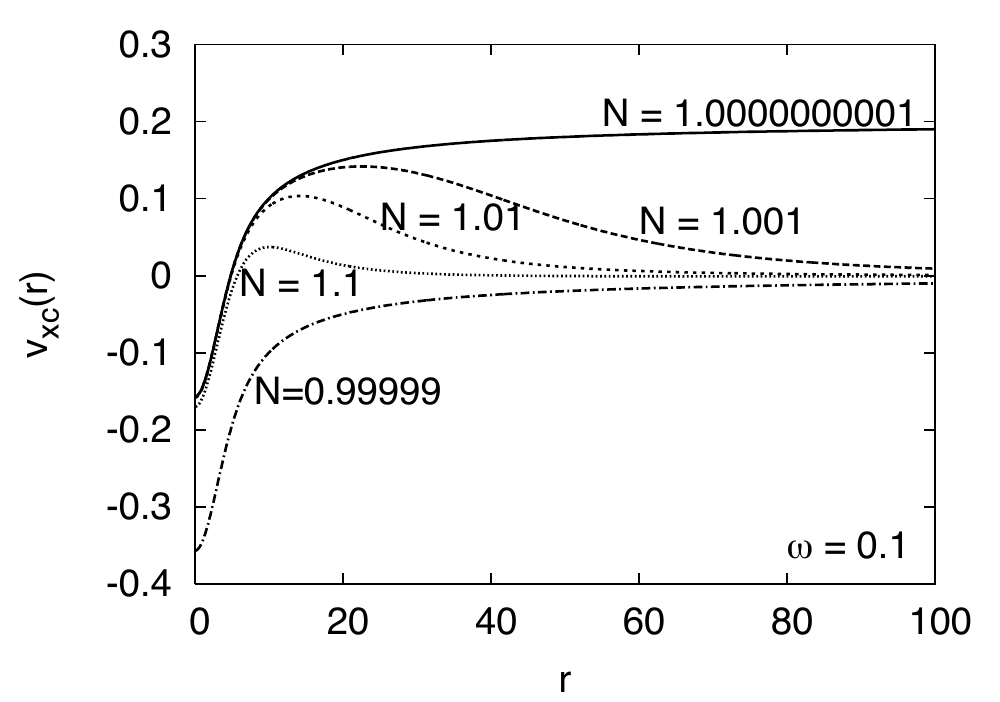}
\includegraphics[width=9cm]{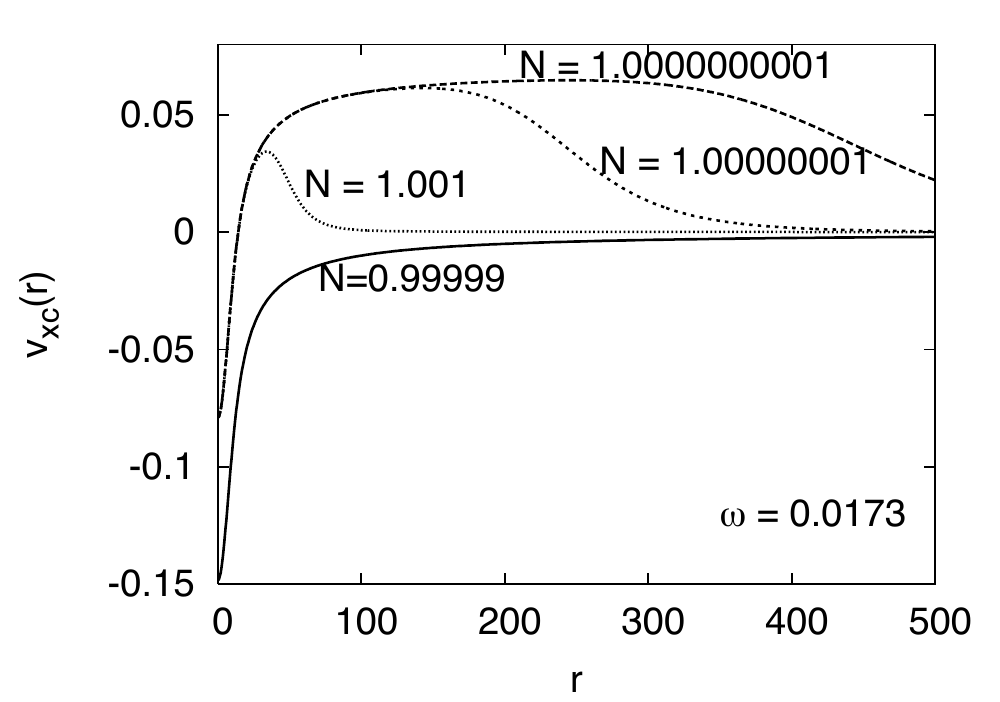}
\caption{The exact exchange-correlation potential for a system with particle number $N$ fluctuating between zero and two in the external potential $\frac{1}{2}\omega r^2$, for three values of $\omega$. The magnitude of the discontinuity at $N=1$ is equal to $E_2-2 E_1$. We have for $\omega=\frac{1}{2}$ $E_2-2 E_1=\frac{1}{2}$, for $\omega=\frac{1}{10}$ $E_2-2 E_1=\frac{1}{5}$, and for $\omega\approx 0.0173$ $E_2-2 E_1=4\, \omega \approx 0.06938$. }
\label{fig_vxc}
\end{figure}

\section{Results}
\label{sec_results}
We considered three cases for which  $\psi(r_{12})$ of Eq.~(\ref{eq_Psi}) is analytical: $\omega=1/2$, $\omega=1/10$, and $\omega=(35-2\sqrt{57})/712\approx 0.0173$ (see Appendix~\ref{app_dens}). The exact exchange-correlation potentials for $N$ slightly below and slightly above 1 are reported in Fig.~\ref{fig_vxc}. The qualitative behavior is similar to the one reported by Sagvolden and Perdew \cite{SagPer-PRA-08} obtained from the model density of the H$^-$ ion: as $N\to 1^+$ the exchange-correlation potential is, over a larger and larger range, more and more equal to the $N\to 1^-$ potential plus a constant. 
The magnitude of the constant is, as expected \cite{PerParLevBal-PRL-82,ShaSch-PRL-83,PerLev-PRL-83}, $I-A$, where $I$ is the ionization potential and $A$ the electron affinity. For closed-shell two-electron systems this amounts to $I-A=E_2-2 E_1$, where $E_2$ is the energy of the system with $N=2$ electrons and $E_1$ is the energy for $N=1$. Figure~\ref{fig_vxc} clearly shows that, as argued in Refs.~\cite{PerParLevBal-PRL-82,ShaSch-PRL-83,PerLev-PRL-83,SagPer-PRA-08}, for any fixed given $N$ (above or below 1)
\beq
\lim_{r \to \infty} v\xc(r)=0,
\eeq
but
\beq
\lim_{r \to \infty} \lim_{N\to 1^+} v\xc(r)=I-A.
\eeq
Notice that when $\omega=1/2$ and $\omega=1/10$ the $N=1.0000000001$ $v_{xc}(r)$ starts to decay to zero for distances much larger than those considered in the figure. The case of the ensemble H and H$^-$ studied by Sagvolden and Perdew \cite{SagPer-PRA-08} resembles most closely to the last case considered here, $\omega\approx 0.0173$. This is due to the fact that the $N=2$ H$^-$ system is considerably correlated. In the case treated here, $v_{\rm ext}(r)=\frac{1}{2}\omega^2 r^2$, the $N=2$ system becomes more and more correlated as $\omega\to 0$. This is also consistent with our previous work \cite{GorSav-JPCS-08}, in which we compared the dependence on $N$ (in the range $1\le N\le 2$) of the non-interacting kinetic energy  of the He series with the one of the Hooke's atom series, finding a resemblence of the H$^-$ case (He series) with the low-$\omega$ case (Hooke's series).

\section{Conclusions}
\label{sec_conc}
The discontinuity of the exact Kohn-Sham exchange-correlation potential was argued in Refs. \cite{PerParLevBal-PRL-82,ShaSch-PRL-83,PerLev-PRL-83}, and is known to have very important physical consequences. In particular, the absence or the underestimation of this discontinuity in local and semilocal functionals explains why they often produce a qualitatively incorrect dissociation limit for non-equilibrium nuclear positions and why they underestimate charge-transfer excitation energies in time-dependent DFT.
In this work we have calculated the exact Kohn-Sham exchange and correlation potentials as the particle number crosses 1 for a system which is analytically soluble. This way, we have illustrated, at least in one case, the existence of this discontinuity without relying on any approximation, thus providing further support to the assumptions that were used for its prediction.

\section*{Acknowledgements} 
It is our pleasure to dedicate this illustration of some simple principles of 
density functional theory to Istvan Mayer who always liked simple theorems, 
proofs, and derivations in quantum chemistry. We are indebted to John P. Perdew for pointing out an error in the previous version, concerning the exchange potential of Eq.~(\ref{eq_vx}). We thank Hector Mera for stimulating discussions and for a critical reading of this manuscript.
This work was supported by the ANR (National French Research Agency) under Grant n.~ANR-07-BLAN-0271.

\appendix
\section{Analytical densities for $N=2$}
\label{app_dens}
As shown by Taut \cite{Tau-PRA-93}, for a set of special values of $\omega$ the ground-state wavefunction of the two-electron hamiltonian has the form
\beq
\Psi(r_1,r_2,r_{12})=\tilde{C}\,e^{-\frac{\omega}{2}(r_1^2+r_2^2)} t(r_{12}),
\eeq
where $r_1=|\rv_1|$, $r_2=|\rv_2|$, $r_{12}=|\rv_2-\rv_1|$, $t(x)$ is a polynomial of finite order $n$ (depending on $\omega$), and $\tilde{C}$ is the normalization constant. The easiest way to obtain the density is
to use the coordinates of Coulson and Nielson \cite{CouNei-PPSL-61}, $r_1$, $r_2$ and $r_{12}$, for the integration over $\rv_2$, 
\beq
\rho_2(r)= C \frac{e^{-\omega\,r^2}}{r}\int_0^\infty d r_{12}\,r_{12}\,t(r_{12})^2\int_{|r_2-r_{12}|}^{|r_2+r_{12}|} dr_2\,r_2\,e^{-\omega\,r_2^2},
\eeq
where $C$ is again a normalization constant. Integration over $r_2$ yields Eq.~(\ref{eq_rho2}).
\subsection{$\omega=1/2$}
The density for $\omega=1/2$ has been reported by several authors (see, e.g., Ref.~\cite{QiaSah-PRA-98}),
\beq
\rho_2(r)=	\frac{e^{-\frac{r^2}{2}}}{4 \left(8+5 \sqrt{\pi }\right) \pi ^{3/2}} \left\{\sqrt{2 \pi } \left[r^2+7+\frac{4
	   \left(r^2+1\right)
	   \text{erf}\left(\frac{r}{\sqrt{2}}\right)}{r}\right]+8
	   e^{-\frac{r^2}{2}}\right\},
\eeq
where $\text{erf}(x)$ is the error function.
\subsection{$\omega=1/10$}
\begin{eqnarray}
\rho_2(r) & = & 	\frac{e^{-\frac{r^2}{5}}}{25000 \pi 
	   \left(61 \pi +48 \sqrt{5 \pi }\right) r} \Biggl\{1000\, r \left(r^2+45\right)+5
	   e^{\frac{r^2}{10}} \sqrt{10 \pi } \Biggl[r \biggl(r^4+190
	   r^2 \nonumber \\
	& + & 2875\biggr)+20 \left(r^4+50 r^2+175\right)
	   \text{erf}\left(\frac{r}{\sqrt{10}}\right)\Biggr]\Biggr\}.
\end{eqnarray}
\subsection{$\omega=(35-2\sqrt{57})/712\approx 0.0173$}
In this case we obtain
\beq
\rho_2(r)=\frac{2}{c}\omega^{3/2}\tilde{\rho}(\sqrt{\omega} \,r),
\eeq
where
\beq
c=\frac{\pi}{384}   \left[3 \left(67741+8855 \sqrt{57}\right) \pi +256
   \left(132+17 \sqrt{57}\right) \sqrt{\left(70+6 \sqrt{57}\right) \pi
   }\right],
\eeq
and
\begin{eqnarray}
\tilde{\rho}(y) & = & \frac{e^{-2 y^2}}{192 \left(-35+3 \sqrt{57}\right)^5 y} \Biggl\{y \Biggl[8 \sqrt{\frac{178}{35-3 \sqrt{57}}}
   \biggl(8 \left(-34410443+4535711 \sqrt{57}\right) y^6 \nonumber \\
& + & 4
   \left(-307865169+39250957 \sqrt{57}\right)
   y^4+\left(1987335802-305824402 \sqrt{57}\right) y^2 \nonumber \\
& + & 1407537905
   \sqrt{57}-11076745221\biggr)+e^{y^2} \sqrt{\pi } \biggl(16
   \left(-34410443+4535711 \sqrt{57}\right) y^8 \nonumber \\
& + & 32
   \left(-394348481+50982845 \sqrt{57}\right) y^6+8
   \left(-773680875+26538047 \sqrt{57}\right) y^4 \nonumber \\ 
& + & 56
   \left(-358975185+8100493 \sqrt{57}\right) y^2+9
   \left(-22356535675+2789329039 \sqrt{57}\right)\biggr)\Biggr] \nonumber \\
& + & 4 e^{y^2}
   \sqrt{\frac{178 \pi }{35-3 \sqrt{57}}} \biggl[16
   \left(-34410443+4535711 \sqrt{57}\right) y^8+32 
   \biggl(-85568903 \nonumber \\
& + & 10946667 \sqrt{57}\biggr) y^6-56
   \left(-53901687+8766619 \sqrt{57}\right) y^4 
 + 72
   \biggl(-272540317 \nonumber \\
& + & 33930025 \sqrt{57}\biggr) y^2+33
   \left(-412010907+53643311 \sqrt{57}\right)\biggr]
   \text{erf}(y)\Biggr\}.
\end{eqnarray}

\end{document}